\documentclass[twocolumn,prd,aps,superscriptaddress,preprintnumbers,tightenlines,showpacs,nofootinbib,eqsecnum,amsfonts,amsmath]{revtex4-1}
\usepackage[normalem]{ulem}
\usepackage{placeins}
\usepackage{color}
\usepackage{calc}
\usepackage{amsmath,amssymb,graphicx}
\usepackage{tensor}
\usepackage{bm}
\usepackage{times}
\usepackage[varg]{txfonts}
\usepackage[colorlinks, pdfborder={0 0 0}]{hyperref}
\usepackage{float}
\usepackage{dcolumn}
\usepackage[nolist,nohyperlinks]{acronym}
\usepackage{xspace}
\usepackage[abs]{overpic}
\usepackage{pict2e}
\usepackage{enumitem}
\usepackage[usenames,dvipsnames]{xcolor}
\usepackage[utf8]{inputenc}
\usepackage{acronym}
\usepackage{gensymb}
\usepackage{cleveref}
\usepackage[normalem]{ulem}
\usepackage{longtable}
\usepackage{verbatim}
\usepackage{multirow}
\usepackage{amsmath,amssymb}
\usepackage{csquotes}

\usepackage{subfigure}
\usepackage{placeins}
\usepackage{hyperref}
\usepackage{aas_macros}
\usepackage{soul}

\definecolor{rb4}{HTML}{27408B}

\begin{document}

\title{Detectability and parameter estimation of GWTC-3 events with LISA }

\author{Alexandre Toubiana}
\affiliation{Max Planck Institute for Gravitational Physics (Albert Einstein Institute), Am Mühlenberg 1, Potsdam-Golm, 14476, Germany}

\author{Stanislav Babak}
\affiliation{Université de Paris, CNRS, Astroparticule et Cosmologie, F-75006 Paris, France}

\author{Sylvain Marsat}
\affiliation{Laboratoire des 2 Infinis - Toulouse (L2IT-IN2P3), Universit\'e de Toulouse, CNRS, UPS, F-31062 Toulouse Cedex 9, France}

\author{Sergei Ossokine}
\affiliation{Max Planck Institute for Gravitational Physics (Albert Einstein Institute), Am Mühlenberg 1, Potsdam-Golm, 14476, Germany}

\begin{abstract}

Multiband observations of coalescing stellar-mass black holes binaries could 
deliver valuable information on the formation of such sources and potential deviations from general relativity. Some of these binaries might be first detected by the space-based detector LISA and, then, several years later, observed with ground-based detectors. Due to large uncertainties in astrophysical models, it is hard to predict the population of such binaries that LISA could observe. In this work, we assess the ability of LISA to detect the events of the third catalog of gravitational wave sources released by the LIGO/Virgo/KAGRA collaboration. We consider the possibility of directly detecting the source with LISA and performing archival searches in the LISA data stream, after the event has been observed with ground-based detectors. We also assess how much could LISA improve the determination of source parameters. We find that it is not guaranteed that any event other than GW150914 would have been detected. Nevertheless, if any event is detected by LISA, even with a very low signal-to-noise ratio, the measurement of source parameters would improve by combining observations of LISA and ground based detectors, in particular for the chirp mass.

\end{abstract}

\maketitle

\section{Introduction}

We are now well into the era of gravitational wave (GW) astronomy. The LIGO-Virgo-KAGRA collaboration (LVK) \cite{LIGOScientific:2014pky,Acernese:2015gua,Aso:2013eba} has recently released its third catalogue of GW sources, GWTC-3, which contains 90 candidate events \cite{LIGOScientific:2021djp}. The increasing number of detections allows us to perform more precise tests of general relativity (GR) \cite{LIGOScientific:2021sio}, to better understand the features of the population of stellar-mass black hole binaries (SBHBs) \cite{LIGOScientific:2021psn} or else to probe the expansion of the Universe \cite{LIGOScientific:2021aug}. The improvement of ground-based detectors over the next few years will provide us with even more numerous and precise detections, increasing the scientific outcome of GW astronomy. In particular, third generation detectors such as the Einstein Telescope (ET) \cite{Punturo:2010zz} or Cosmic Explorer (CE) \cite{Reitze:2019iox} will observe virtually all merging SBHBs in the Universe, some of them with a signal-to-noise ratio (SNR) up to thousands.

Scheduled for 2034, the Laser Interferometer Space Antenna \cite{Audley:2017drz} will observe GWs in the mHz band and should be able to detect and resolve the loudest SBHBs during their early inspiral, allowing for {\it multiband} observations \cite{Sesana:2016ljz}. Though the main targets of the LISA mission are massive black hole binaries \cite{Klein:2015hvg} and galactic binaries \cite{Korol:2017qcx}, the scientific potential of SBHB observations with LISA is enormous. Such observations could be used to detect low frequency modifications to the signal, caused by deviations from GR \cite{Toubiana:2020vtf,Gupta:2020lxa,Perkins:2020tra,Gnocchi:2019jzp,Carson:2019rda,Chamberlain:2017fjl,Vitale:2016rfr,Barausse:2016eii} and/or astrophysical effects \cite{Toubiana:2020drf,Caputo:2020irr,Tamanini:2019usx,Cardoso:2019rou,Barausse:2014tra,Barausse:2014pra}, or to inform ground-based detectors and allow them to improve their sensitivity to the merger of the system \cite{Gerosa:2019dbe,Tso:2018pdv}, allowing for more precise ringdown tests. Moreover, the precise sky localization provided by LISA could be used to inform electromagnetic instruments about a potential counterpart \cite{Toubiana:2020cqv,Toubiana:2020drf,Caputo:2020irr}, simplifying multimessenger observations, and to tighten constraints on the Hubble constant, even in the case of dark sirens \cite{Muttoni:2021veo}. Finally, the orbit of SBHBs is circularized to a high degree for frequencies accessible to ground-based detectors \cite{Peters:1964zz}, whereas the orbital eccentricity could be substantial in the LISA frequency band. By measuring the eccentricity, LISA could help to distinguish between different formation channels \cite{Klein:2022rbf,Buscicchio:2021dph,Samsing:2018isx,Nishizawa:2016jji,Nishizawa:2016eza,Breivik:2016ddj}.

The number and properties of SBHBs that LISA will detect are hard to predict, due to the large uncertainties in the population of these systems. Therefore, we adopt a data-driven approach and investigate the ability of LISA to detect the systems reported by the LVK so far and the improvement in their parameter estimation thanks to LISA. We consider both direct observations, where LISA would detect a system before ground-based detectors, and ``archival'' ones, where we use the information provided by ground-based detectors to reduce the parameter space and facilitate the detection of those sources in the LISA data \cite{Wong:2018uwb}. On one hand, we find that even in the most optimistic configurations only a few sources in GWTC-3 could have been detected by LISA. On the other hand, we show that even for systems with low SNR in LISA, the large number of cycles in its sensitivity band would improve the parameter estimation relative to ground-based detectors alone, even for ET. Our study gives a realistic overview of the potential of SBHBs detections with LISA, though more massive systems, to which ground-based detectors are less sensitive, could be more easily detected with LISA. 

This paper is organized as follows: in Sec.~\ref{sec:signal} we review the tools of LISA data analysis that we use in this work, in Sec.~\ref{sec:catalogue} we describe how do we define LISA events from GWTC-3, in Sec.~\ref{sec:results} we present our results for the detectability of these events and parameter estimation for some selected events, and in Sec.~\ref{sec:ccl} comment on those results and draw conclusions.

\section{Generation of LISA signals and data analysis}\label{sec:signal}

We briefly summarize our tools for the analysis of SBHBs with LISA, and refer to our thorough investigation of parameter estimation for SBHBS with LISA in \cite{Toubiana:2020cqv}. 

Since LISA will observe SBHBs during their early inspiral, we consider only the dominant $\ell=2,m=2$ mode and neglect the contribution of higher modes. Moreover, since during the early inspiral precession effects are expected to be subdominant, we consider only spins components parallel to the orbital angular momentum. Note that precession effects would be important when relating the observations in the LISA band to the ones in ground-based detectors band, as in the case of a full multiband study. Finally, due to the limitation of our waveform model, we only consider quasicircular binaries. Eccentricity information was not provided in GWTC-3, the signals being compatible with circularized orbits in the LIGO/Virgo band, but this is in principle an important limitation of our current analysis, as eccentricity might be important in the LISA band. In particular, based on the results of \cite{Klein:2022rbf,Buscicchio:2021dph}, we expect the estimation of the time to coalescence and of the chirp mass to be much and slightly more optimistic respectively. A signal is then described by 11 parameters: the individual (detector-frame) masses $m_1$ and $m_2$ and dimensionless spins $\chi_1$ and $\chi_2$, the initial frequency of the signal in the LISA band $f_0$, the luminosity distance $D_L$, the inclination $\iota$, the longitude $\lambda$ and the polar angle $\beta$ in the solar system barycenter frame, the polarization $\psi$ and the initial phase $\varphi_0$. We denote this set of parameters by $\theta$. We generate the GW signal using the phenomenological waveform PhenomD \cite{Husa:2015iqa,Khan:2015jqa} and compute the full LISA response as described in \cite{Marsat:2018oam, Marsat:2020rtl} to obtain the three time-delay interferometry (TDI) \cite{Tinto:2004wu} channels $A$, $E$ and $T$. They constitute three noise-uncorrelated data streams, the equivalent of the measured strain in ground-based detectors.

In general, the observed data stream in each of the TDI channels (labeled by $c$) will be a superposition of GW signals $s_{i,c}$ and noise: $d_c=\sum_i s_{i,c}+n_c$. Here, we work in the zero-noise approximation ($n=0$), which can be seen as an average over noise realizations \cite{Rodriguez:2013oaa}, and consider only one GW signal at a time. These slowly evolving but nonmonochromatic signals have a very distinct morphology from the other LISA sources, such as massive black hole binaries and galactic binaries, and, given the small expected number of resolvable events (as we will show in this paper), it is unlikely that they overlap in the same time-frequency bins. Therefore, it is reasonable to treat these signals individually. We neglect systematic errors due to modeling, and consider that the GW signal emitted by the source and observed with LISA in each TDI channel can be represented by a GW template $s(\theta)$.

We define the noise-weighted inner product between two data streams as 
\begin{equation}
    (h_1|h_2)=4 {\mathcal Re} \left ( \int_0^{+\infty}\frac{h_1(f)h^*_2(f)}{S_n(f)} \right ),
\end{equation}
where $S_n(f)$ is the power spectral density (PSD). In this work, we use the SciRDv1 noise curve \cite{scirdv1}, which sets pessimistic limits on the noise level, as well as the ``current best estimate'' (CBE), for which the level of noise is about a factor 1.5 lower across the LISA frequency band occupied by SBHBs. For a given choice of the PSD, the SNR of a signal $h$ is defined as ${\rm SNR}=\sqrt{(h|h)}$. 

To estimate the parameters of detectable systems, we work in a Bayesian framework and compute their posterior distribution given an observed dataset $d$ using Bayes' theorem:
\begin{equation}
    p(\theta|d)=\frac{p(d|\theta)p(\theta)}{p(d)},
\end{equation}
where $p(d|\theta)$ is the likelihood, $p(\theta)$ is the prior and $p(d)$ is the evidence. As long as we are not interested in model selection, the latter acts as a normalization constant, and thus can be discarded. We take the prior to be flat in individual masses and spins, volume (as $D_L^2$) and flat in $\cos(\iota)$, as well as in the remaining parameters, as the \emph{Fiducial} prior in \cite{Toubiana:2020cqv}. Under the hypothesis that noise is stationary and Gaussian the likelihood is given by:
\begin{equation}
    p(d|\theta) \propto \prod_{c \in [A,E,T]}\exp \left( -\frac{1}{2}(d_c-s_c(\theta)|d_c-s_c(\theta)) \right ).
\end{equation}
The posterior distribution is then sampled via a Markov chain Monte Carlo algorithm (MCMC), the same one introduced in \cite{Toubiana:2020cqv}. In particular, we use the chirp mass, $\mathcal{M}_c=\left ( \frac{m_1^3m_2^3}{(m_1+m_2)} \right )^{(1/5)}$, the symmetric mass ratio, $\eta=\frac{m_1m_2}{(m_1+m_2)^2}$, and the mass-weighted symmetric and antisymmetric spin combinations $\pmb{ \chi_{\pm}=\frac{m_1\chi_1\pm m_2\chi_2}{m_1+m_2}}$, instead of the individual masses and spins for the sampling. $\chi_+$ is often called the effective spin. 

\section{From GWTC-3 to LISA events}\label{sec:catalogue}

 \begin{figure*}[hbtp!]
  \includegraphics[scale=0.37]{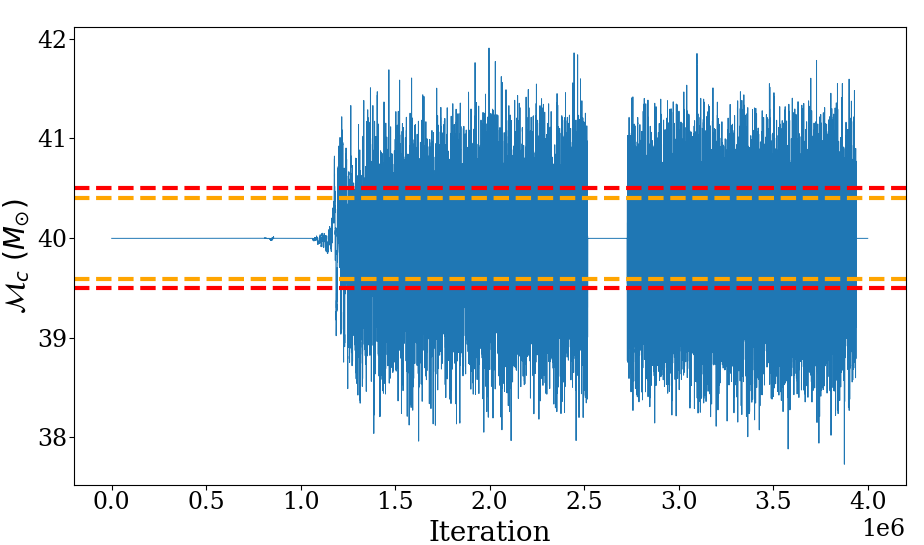}
    \includegraphics[scale=0.37]{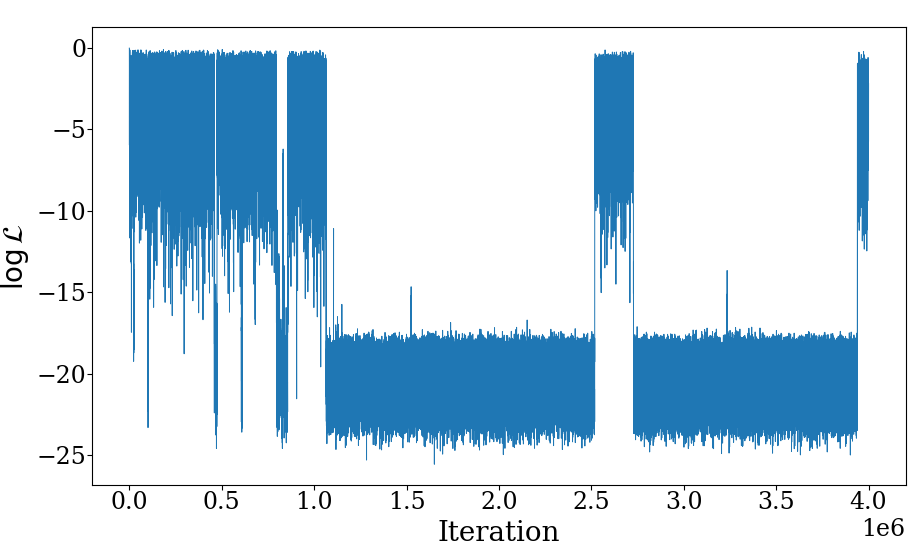}
   \caption{MCMC chain of the chirp mass and the log-likelihood for a system with LISA SNR 4.5 and ground-based SNR of 240. The drop in likelihood for several iterations and the width of the chirp mass region explored show that the sampler has \enquote{lost} the signal and is exploring the prior. Because in this case we use a Gaussian prior centered at the injection point, the portion of the parameter space that the chain can explore is very limited, and it ends up finding the injected signal again. This would most likely not happen if we were to use a more agnostic prior.} The red dashed lines indicate the $1 \sigma$ interval around the true value of the Gaussian prior used to mimic ET, and the orange ones the 1$\sigma$ interval of the chain itself. The latter is slightly narrower than the former, due to the non-negligible fraction of points that actually explore the posterior. Note that what appears as a line (source sampling) actually has a finite width which is 4 orders of magnitude thinner than the prior (wide regions). This shows that systems at such low SNR might not be confidently detected by LISA, even in the case of archival searches following ET detections with a reduced prior.\label{fig:mcmc_chain}  
 \end{figure*}

For the exploration here we consider all events as detected by the LVK in all observing runs up to and including O3b. We use the latest parameter estimation results from GWTC-2.1 \cite{LIGOScientific:2021usb} for all events in O1, O2 and O3a, and results from GWTC-3\cite{LIGOScientific:2021djp} for events in O3b. We use samples which have been reweighted to a prior which is uniform in comoving volume. To minimize the impact of waveform systematics, we use the `C01:Mixed` datasets for all cases\footnote{Except for the two BNS events.}, which include a mixture of two precessing, higher-order mode models \texttt {IMRPhenomXPHM}\cite{Pratten:2020ceb} and \texttt{SEOBNRv4PHM}\cite{Ossokine:2020kjp} data for most events (see Ref.~\cite{LIGOScientific:2021djp} for more details). Note that the prior used by the LVK for distance (uniform in comoving volume) and spin (uniform in magnitude and orientation for each 3-dimensional spin) differs from ours. Moreover, we use the flat prior on $f_0$ that does not translate into a flat prior for $t_c$, as the one used by the LVK. This is not an issue since here we use the LVK results as a source list and use their constraints on the sources' parameters.

For each candidate event in GWTC-3 we use the samples for $\mathcal{M}_c$, $\eta$, $\chi_+$, $\chi_-$, $\cos(\iota)$ and $\log_{10}(D_L)$. We take the aligned component of spins by projecting them on the orbital angular momentum. We discard the information about sky position, polarization and phase because we take those sources as representative, while these parameters could be anything during the LISA's mission time. Therefore, we draw randomly these parameters for each sample (uniform on the sphere for the sky location and uniform in $[-\pi,\pi]$ for polarization and phase). We retain the inclination information provided by the LVK because it is strongly correlated with the distance, for which the distribution over binaries should depend on their formation channels. As extensively discussed in \cite{Toubiana:2020cqv}, the initial GW frequency (or orbital separation) at the beginning of the LISA observation has a great impact on its detectability and subsequent parameter estimation. The best results are achieved when the binary is observed as close to merger as possible, i.e.~when it starts chirping. Therefore, in order to span the range of possibilities, we consider three combinations of the time to coalescence ($t_c$) and the observation time with LISA ($T_{\rm obs}$), ranging from most pessimistic to most optimistic: ($t_c=10{\rm yr}$, $T_{\rm obs}=6{\rm yr}$), ($t_c=6{\rm yr}$, $T_{\rm obs}=6{\rm yr}$) and ($t_c=10{\rm yr}$, $T_{\rm obs}=10{\rm yr}$). In the first case, the LISA observations end before the binary has merged, while in the two latter cases, we observe the chirp toward the merger as it happens in LISA.

Following the above procedure to define LISA events, we compute their SNR for each $t_c$ and $T_{\rm obs}$ combination, using both the SciRDv1 and CBE noise curves. Finally, we multiply the computed SNR by $\sqrt{0.75}$ to account for the estimated $75\%$ duty cycle of LISA. This is a good approximation for almost monochromatic signals with mildly changing amplitude during the observational period, and it is an acceptable approximation for SBHBs. However, this simple scaling cannot be applied when performing parameter estimation, since a lot of the constraining power comes from the late inspiral, more than the relative contribution of the late inspiral to the total SNR \cite{Toubiana:2020cqv}. Though the search for SBHB signals in the LISA data stream is known to be non-trivial \cite{Moore:2019pke}, for simplicity we will assume that a signal is detectable by LISA alone if its SNR is above 8, and with the help of ground-based detectors, i.e.~performing archival searches, if its SNR is above 5. We will somewhat justify these choices later in Sec.~\ref{sec:results}. We then perform parameter estimation for a GW191109$\_$010717-like (source detectable by LISA alone ) and a GW200112$\_$155838-like (source detectable through the archival search) systems. We pessimistically use SciRDv1 noise curve when performing parameter estimation, but do not account for the effect of duty cycle.   

When performing parameter estimation on selected events, we mimic a multiband detection with LISA and LVK-like detectors by fixing the time to coalescence, since it is very accurately measured by ground-based detectors, and using a Gaussian prior \footnote{Unlike some events in GWTC-3, the two events for which we perform parameter estimation here do not present multimodality in the parameters listed above, and exhibit only a mild departure from gaussianity. Therefore, this approximation should not qualitatively change the results.} with covariance matrix computed from the LVK samples for $\mathcal{M}_c$, $\eta$, $\chi_+$, $\chi_-$, $\cos(\iota)$ and $\log_{10}(D_L)$\footnote{ Our prior when mimicking multiband detections is no longer the \emph{Fiducial} one described in Sec.~\ref{sec:signal}.}. We center the Gaussian prior at the injection point. We call this scenario LISA+LVK. When fixing the time to coalescence, the initial frequency $f_0$ is no longer left free to vary, and it is computed from the values of masses and spins such that the coalescence time is the one we have fixed. In practice, we apply the stationary phase approximation to the full PhenomD phase to get the time-frequency relation, and invert it numerically to compute the initial frequency. Neglecting eccentricity and precession leads to a tighter inference of $f_0$. We also consider a LISA+ET$_r$ scenario, where $r$ stands for \enquote{rescaled}, by dividing the elements of the prior covariance matrix by $16^2$. The factor $16$ corresponds to the average improvement in the SNR from LVK-like detectors to ET for the systems in GWTC-3, as we have computed using the ET noise curve EinsteinTelescopeP1600143 implemented in pycbc \cite{alex_nitz_2019_3483184}. We then use the standard approximation that measurement errors scale as $1/{\rm SNR}$. Notice that our LISA+ET$_r$ scenario is most likely \footnote{The uncertainty resides in the fact that starting from rather low SNR systems, for which the 1/SNR scaling is not expected to apply, it is hard to estimate if this approximation overestimates or underestimates how the measurement improves with the increase in SNR.} conservative, since in addition to having a lower noise level, ET would observe more cycles of SBHB systems. This would, in particular, contribute to improve the estimation of the chirp mass.

\section{Results}\label{sec:results}

\begin{figure*}[hbtp!]
  \includegraphics[scale=0.25]{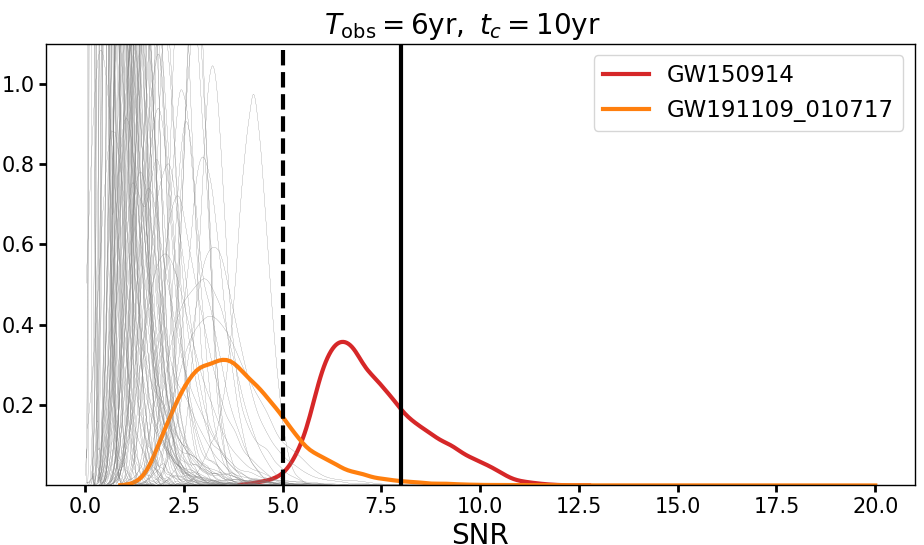}
    \includegraphics[scale=0.25]{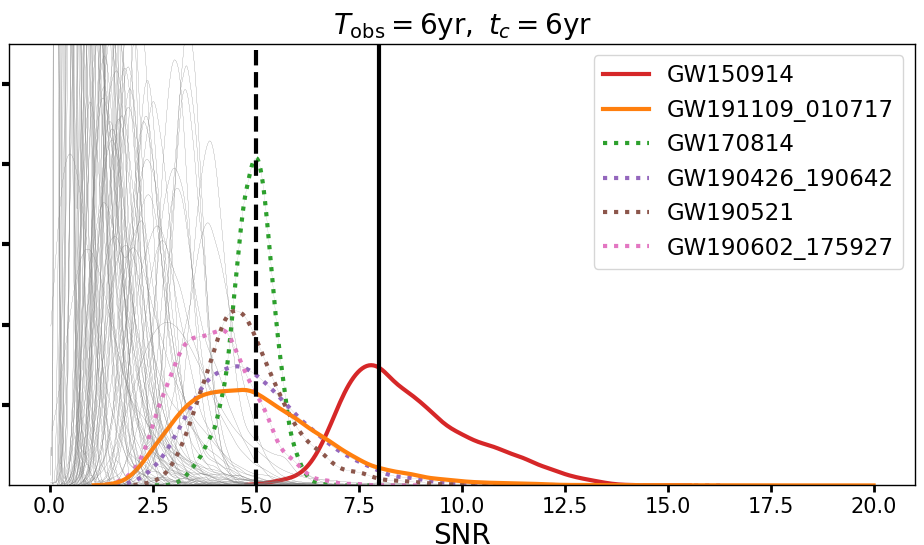}
    \includegraphics[scale=0.25]{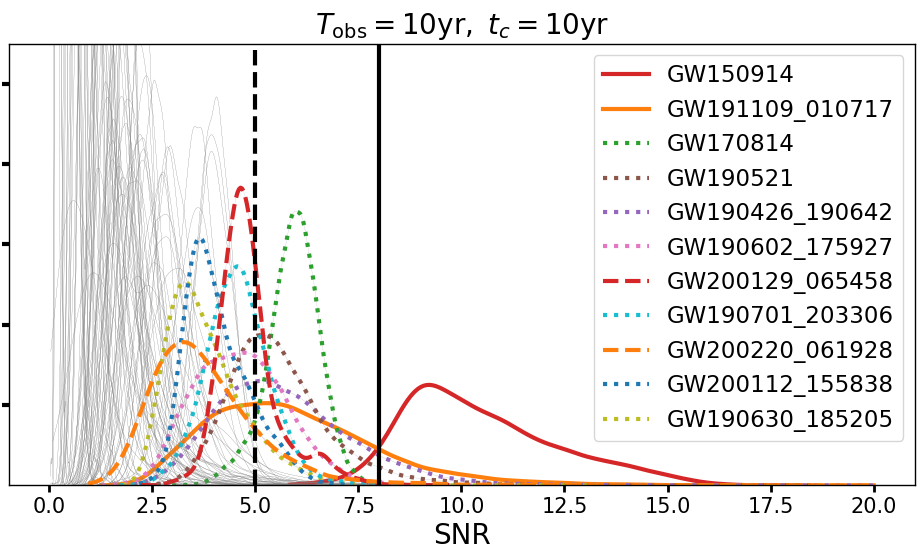}
   \caption{Distribution of SNRs in LISA of events in GWTC-3 for different configurations, using the CBE PSD. As expected from our previous study \cite{Toubiana:2020cqv}, observing the sources closer to merger increases significantly the probability of detection. The black solid (dashed) line indicates the threshold for direct (archival) detection. We show in color the events with $p_{\rm det}^5>0.1$ (defined in the main text). We use solid lines to differentiate GW150914 and GW191109$\_$010717, because they are the only events that have $p_{\rm det}^8>0.1$ in the $t_c=10{\rm yr}$, $T_{\rm obs}=10{\rm yr}$ configuration, indicating that they are the only ones directly detectable by LISA.} GW150914 and GW191109$\_$010717 are shown in solid lines because they are the only events that have $p_{\rm det}^8>0.1$ in the $t_c=10{\rm yr}$, $T_{\rm obs}=10{\rm yr}$ configuration, indicating that they are the only directly detectable by LISA. \label{fig:snrs}   
 \end{figure*}

 To ease the comprehension of the text, we provide a short summary of the working hypotheses used in the following sections.

\begin{itemize}
    \item In Sec.~\ref{subsec:det}, we assume a LISA duty cycle of $75\%$ and consider both the SciRDv1 and the CBE noise curves. We report detection probabilities (as defined in that section) computed with each noise curve in Table \ref{tab:pdets}. Fig.~\ref{fig:snrs} shows SNRs computed with the CBE noise curve.
    \item For our parameter estimation runs in Sec.~\ref{subsec:pe}, we do not consider the effect of duty cycle, and use the SciRDV1 noise curve. In the LISA-only runs we use the \emph{Fiducial} prior defined in Sec.~\ref{sec:signal}, whereas when mimicking multiband runs we use a Gaussian prior, as described in Sec.~\ref{sec:catalogue}. 
    \item For the three combinations of $t_c$ and $T_{\rm obs}$ that we consider, we define two SNR thresholds: 8 for a direct detection with LISA and 5 for an archival detection. In Sec.~\ref{subsec:det} we focus on the LVK events that have an archival detection probability larger than 0.1 in the most optimistic case ($t_c=T_{\rm obs}=10{\rm yr}$). 
    \end{itemize}

We start by discussing our choice of SNR threshold for detection with LISA.

\begin{table*}
 \begin{center}
   \begin{tabular}{c *{4}{c|}}
   \cline{3-5}

  & & \multicolumn{1}{|c|}{$t_c=10{\rm yr}$, $T_{\rm obs}=6{\rm yr}$} & \multicolumn{1}{|c|}{$t_c=6{\rm yr}$, $T_{\rm obs}=6{\rm yr}$} & \multicolumn{1}{|c|}{$t_c=10{\rm yr}$, $T_{\rm obs}=10{\rm yr}$} \\

\hline

  \multicolumn{1}{|c|}{\multirow{2}{*}{GW150914}} &  \multicolumn{1}{|c|}{$p_{\rm det}^5$} & 
  \multicolumn{1}{|c|}{$7.1 $ - $9.9 \times 10^{-1}$} & \multicolumn{1}{|c|}{$0.9$ - $1$}& 
  \multicolumn{1}{|c|}{$1$ } \\

   \multicolumn{1}{|c|}{}& \multicolumn{1}{|c|}{$p_{\rm det}^8$} & \multicolumn{1}{|c|}{ $0.3$ - $1.9 \times 10^{-1}$} & \multicolumn{1}{|c|}{$1.6$ - $4.6 \times 10^{-1}$}& \multicolumn{1}{|c|}{ $4.2$ - $9.1 \times 10^{-1}$} \\
  
  \hline

  \multicolumn{1}{|c|}{\multirow{2}{*}{GW170814}} & \multicolumn{1}{|c|}{$p_{\rm det}^5$} &
  \multicolumn{1}{|c|}{$0.02$ - $2.1 \times 10^{-2}$} & 
  \multicolumn{1}{|c|}{$0.07$ - $3.3 \times 10^{-1}$ }& \multicolumn{1}{|c|}{$2.7$ - $8.7 \times 10^{-1}$ } \\

  \multicolumn{1}{|c|}{} &  \multicolumn{1}{|c|}{$p_{\rm det}^8$} & \multicolumn{1}{|c|}{$0$} & 
  \multicolumn{1}{|c|}{$0$-$3.6 \times 10^{-5}$}& 
  \multicolumn{1}{|c|}{$0$ - $1.1 \times 10^{-3}$ } \\
  
  \hline
    
    \multicolumn{1}{|c|}{\multirow{2}{*}{GW190426$\_$19064}} & \multicolumn{1}{|c|}{$p_{\rm det}^5$} & 
    \multicolumn{1}{|c|}{ $0.8 $ - $4.8 \times 10^{-2}$ } & \multicolumn{1}{|c|}{$1.4$ - $4.0 \times 10^{-1}$ }& \multicolumn{1}{|c|}{$2.9$ - $5.9 \times 10^{-1}$ } \\

   \multicolumn{1}{|c|}{}& \multicolumn{1}{|c|}{$p_{\rm det}^8$} & \multicolumn{1}{|c|}{$2.4$ - $7.7 \times 10^{-4}$} & \multicolumn{1}{|c|}{$0.3$ - $1.9 \times 10^{-2}$}& \multicolumn{1}{|c|}{$1.0$ - $5.6 \times 10^{-2}$ } \\
  
  \hline
    
    \multicolumn{1}{|c|}{\multirow{2}{*}{GW190521$\_$030229}} & \multicolumn{1}{|c|}{$p_{\rm det}^5$} &
    \multicolumn{1}{|c|}{ $1.4$ - $4.5 \times 10^{-2}$} & \multicolumn{1}{|c|}{$0.9$ - $2.8 \times 10^{-1}$ }& \multicolumn{1}{|c|}{$2.0$ - $5.2 \times 10^{-1}$} \\

  \multicolumn{1}{|c|}{} &\multicolumn{1}{|c|}{$p_{\rm det}^8$}  & \multicolumn{1}{|c|}{ $0.3$ - $2.1 \times 10^{-3}$} & \multicolumn{1}{|c|}{$0.6$ - $2.0 \times 10^{-2}$}& \multicolumn{1}{|c|}{$1.4$ - $4.3 \times 10^{-2}$ } \\
  
  \hline

    \multicolumn{1}{|c|}{\multirow{2}{*}{GW190521$\_$074359}} & \multicolumn{1}{|c|}{$p_{\rm det}^5$} & 
    \multicolumn{1}{|c|}{ $1.0$ - $3.1 \times 10^{-2}$} & \multicolumn{1}{|c|}{$3.3$ - $9.7 \times 10^{-2}$ }& \multicolumn{1}{|c|}{$0.8$ - $2.6 \times 10^{-1}$} \\

  \multicolumn{1}{|c|}{} &\multicolumn{1}{|c|}{$p_{\rm det}^8$}  & \multicolumn{1}{|c|}{ $0$ - $1.8 \times 10^{-3}$} & \multicolumn{1}{|c|}{$1.5$ - $6.2 \times 10^{-3}$}& \multicolumn{1}{|c|}{$0.5$ - $2.0 \times 10^{-2}$ } \\
  
  \hline
    
    \multicolumn{1}{|c|}{\multirow{2}{*}{GW190602$\_$175927}} & \multicolumn{1}{|c|}{$p_{\rm det}^5$} & 
    \multicolumn{1}{|c|}{ $1.3$ - $8.9 \times 10^{-3}$} & \multicolumn{1}{|c|}{$0.2$ - $1.3 \times 10^{-1}$ }& \multicolumn{1}{|c|}{$0.9$ - $3.4 \times 10^{-1}$ } \\

   \multicolumn{1}{|c|}{}  & \multicolumn{1}{|c|}{$p_{\rm det}^8$} & \multicolumn{1}{|c|}{$0$-$5.8 \times 10^{-5}$} & 
   \multicolumn{1}{|c|}{$0.1$ - $1.4 \times 10^{-3}$}& \multicolumn{1}{|c|}{$1.0$ - $5.9 \times 10^{-3}$} \\
  
  \hline
  
    \multicolumn{1}{|c|}{\multirow{2}{*}{GW190630$\_$185205}} & \multicolumn{1}{|c|}{$p_{\rm det}^5$} &
    \multicolumn{1}{|c|}{ $0.3 $ - $2.8 \times 10^{-3}$ } & \multicolumn{1}{|c|}{$0.2$ - $2.3 \times 10^{-2}$ }& \multicolumn{1}{|c|}{$0.2$ - $1.1 \times 10^{-1}$} \\

   \multicolumn{1}{|c|}{} & \multicolumn{1}{|c|}{$p_{\rm det}^8$} & \multicolumn{1}{|c|}{$0$} & 
   \multicolumn{1}{|c|}{$0$-$6.0 \times 10^{-5}$}& 
   \multicolumn{1}{|c|}{ $0.6$ - $3.6 \times 10^{-4}$} \\
  
  \hline

\multicolumn{1}{|c|}{\multirow{2}{*}{GW190701$\_$203306}} & \multicolumn{1}{|c|}{$p_{\rm det}^5$} & 
\multicolumn{1}{|c|}{ $0.1 $ - $1.2 \times 10^{-3}$} & 
\multicolumn{1}{|c|}{$0.2$ - $3.2 \times 10^{-2}$}& \multicolumn{1}{|c|}{$0.2$ - $2.3 \times 10^{-1}$} \\

 \multicolumn{1}{|c|}{} & \multicolumn{1}{|c|}{$p_{\rm det}^8$} & \multicolumn{1}{|c|}{$0$} & 
 \multicolumn{1}{|c|}{$0$ - $8.9 \times 10^{-5}$}& 
 \multicolumn{1}{|c|}{ $0.9$ - $6.6 \times 10^{-4}$} \\
  
  \hline

  \multicolumn{1}{|c|}{\multirow{2}{*}{GW190706$\_$222641}} & \multicolumn{1}{|c|}{$p_{\rm det}^5$} &
  \multicolumn{1}{|c|}{ $0.7 $ - $2.5 \times 10^{-2}$} & \multicolumn{1}{|c|}{$4.0$ - $9.5 \times 10^{-2}$}& \multicolumn{1}{|c|}{$0.8$ - $1.5 \times 10^{-1}$} \\

 \multicolumn{1}{|c|}{} & \multicolumn{1}{|c|}{$p_{\rm det}^8$} & \multicolumn{1}{|c|}{$0.5$-$1.9 \times 10^{-3}$} & \multicolumn{1}{|c|}{$3.2$ - $8.1 \times 10^{-3}$}& 
 \multicolumn{1}{|c|}{ $0.7$ - $2.0 \times 10^{-2}$} \\
  
  \hline

\multicolumn{1}{|c|}{\multirow{2}{*}{GW191109$\_$010717}} & \multicolumn{1}{|c|}{$p_{\rm det}^5$} & 
\multicolumn{1}{|c|}{ $0.7 $ - $1.9 \times 10^{-1}$} & \multicolumn{1}{|c|}{$2.4$ - $4.5 \times 10^{-1}$}& \multicolumn{1}{|c|}{$4.0$ - $6.3 \times 10^{-1}$} \\

  \multicolumn{1}{|c|}{} & \multicolumn{1}{|c|}{$p_{\rm det}^8$} & \multicolumn{1}{|c|}{$0.3$ - $1.3 \times 10^{-2}$} & \multicolumn{1}{|c|}{$1.9$ - $6.5 \times 10^{-2}$}& \multicolumn{1}{|c|}{$0.5$ - $1.4 \times 10^{-1}$} \\
  
  \hline
  
  \multicolumn{1}{|c|}{\multirow{2}{*}{GW200112$\_$155838}} & \multicolumn{1}{|c|}{$p_{\rm det}^5$} & 
  \multicolumn{1}{|c|}{$2.3$-$6.9 \times 10^{-4}$ } & \multicolumn{1}{|c|}{$0.09$ - $1.7 \times 10^{-2}$}& \multicolumn{1}{|c|}{$1.1$ - $1.2 \times 10^{-1}$} \\

  \multicolumn{1}{|c|}{} & \multicolumn{1}{|c|}{$p_{\rm det}^8$} & \multicolumn{1}{|c|}{$0$ } &
  \multicolumn{1}{|c|}{ $0$}& 
  \multicolumn{1}{|c|}{$0$-$2.3 \times 10^{-4}$} \\
  
  \hline

\multicolumn{1}{|c|}{\multirow{2}{*}{GW200129$\_$065458}} & \multicolumn{1}{|c|}{$p_{\rm det}^5$} & 
\multicolumn{1}{|c|}{ $0.05$ - $1.6 \times 10^{-2}$} & \multicolumn{1}{|c|}{$0.9$ - $8.5 \times 10^{-2}$ }& \multicolumn{1}{|c|}{$0.7$ - $3.2 \times 10^{-1}$} \\

  \multicolumn{1}{|c|}{} & \multicolumn{1}{|c|}{$p_{\rm det}^8$} & \multicolumn{1}{|c|}{$0$} & 
  \multicolumn{1}{|c|}{$0$}&
  \multicolumn{1}{|c|}{ $0$ - $3.0 \times 10^{-3}$} \\
  
  \hline
  
  \multicolumn{1}{|c|}{\multirow{2}{*}{GW200220$\_$061928}} & \multicolumn{1}{|c|}{$p_{\rm det}^5$} &
  \multicolumn{1}{|c|}{ $0.8 $ - $1.5 \times 10^{-2}$ } &
  \multicolumn{1}{|c|}{$2.4$ - $8.4 \times 10^{-2}$ }& \multicolumn{1}{|c|}{$0.5$ - $1.5 \times 10^{-1}$} \\

   \multicolumn{1}{|c|}{} &  \multicolumn{1}{|c|}{$p_{\rm det}^8$} & \multicolumn{1}{|c|}{$2.0$ - $3.6 \times 10^{-3}$} & \multicolumn{1}{|c|}{$0.5$ - $7.1 \times 10^{-2}$}& \multicolumn{1}{|c|}{$0.8$ - $1.4 \times 10^{-2}$} \\
  
  \hline

   \multicolumn{1}{|c|}{\multirow{2}{*}{GWTC-3}} & \multicolumn{1}{|c|}{$p_{\rm det}^5$} &
   \multicolumn{1}{|c|}{ $7.5 $ - $9.9 \times 10^{-1}$ } & \multicolumn{1}{|c|}{$0.9$ - $1$ }&
   \multicolumn{1}{|c|}{$1$} \\

   \multicolumn{1}{|c|}{} &  \multicolumn{1}{|c|}{$p_{\rm det}^8$} & \multicolumn{1}{|c|}{$0.4$ - $2.1 \times 10^{-1}$} & \multicolumn{1}{|c|}{$2.0$ - $5.3 \times 10^{-1}$}& \multicolumn{1}{|c|}{$4.8$ - $9.3 \times 10^{-1}$} \\
  
  \hline

   \multicolumn{1}{|c|}{\multirow{2}{*}{GWTC-3 - GW150914 }} & \multicolumn{1}{|c|}{$p_{\rm det}^5$} & 
   \multicolumn{1}{|c|}{ $1.2 $ - $3.7 \times 10^{-1}$ } & \multicolumn{1}{|c|}{$5.0$ - $9.2 \times 10^{-1}$ }& \multicolumn{1}{|c|}{$8.6$ - $9.9 \times 10^{-1}$} \\

   \multicolumn{1}{|c|}{} &  \multicolumn{1}{|c|}{$p_{\rm det}^8$} & \multicolumn{1}{|c|}{$0.4$ - $2.6 \times 10^{-2}$} & \multicolumn{1}{|c|}{$0.4$ - $1.4 \times 10^{-1}$}& \multicolumn{1}{|c|}{$1.0$ - $2.9 \times 10^{-1}$} \\
  
  \hline

   \end{tabular}
\end{center}
 \caption{Detection probabilities for the three $t_c$ and $T_{\rm obs}$ combinations. The lower value is obtained assuming the SciRDv1 noise curve and the higher value, the CBE one. Both account for the $75\%$ duty cycle. We give the values only for the 13 events that have $p_{\rm det}^5>0.1$ in the most favorable case ($t_c=10{\rm yr}$, $T_{\rm obs}=10{\rm yr}$, CBE PSD). If instead we use the SciRDv1 noise curve, only 4 events pass that criteria, and only GW150914 if we further take $t_c=10{\rm yr}$ and $T_{\rm obs}=10{\rm yr}$. The last two rows indicate the probability of detecting at least one event in GWTC-3, with and without GW150914, assuming all the events are independent.}\label{tab:pdets}
\end{table*}

\subsection{SNR threshold}

In this study we do not perform a proper search for signals in the LISA data stream, our MCMC chains start close to the injection. Despite that, sometimes the chains \enquote{lose} the signal, depending on the SNR of the system in LISA and the width of the prior. This is illustrated in Fig.~\ref{fig:mcmc_chain}. There, we can clearly see that the drop in log-likelihood is associated with the chain exploring the (Gaussian) prior for the chirp mass. More generally, the low-likelihood regions correspond to the chain exploring the prior for all the parameters of the system. The system for which we run our MCMC has a LISA SNR of 4.5, and we have rescaled the covariance matrix used in the Gaussian prior such that it corresponds to an SNR of 240 in ground-based detectors, a typical value for ET. Varying the SNR of the system in LISA, our MCMC code unambiguously \enquote{finds} the signal for LISA SNRs above $\sim 5$ when using a similar Gaussian prior, and above $\sim 8$ when using uniform (LISA only) prior. This behavior justifies our choice of thresholds for archival and direct detection of SBHBs with LISA. Note that our threshold for archival detection is higher than the value 4 that was originally suggested in \cite{Wong:2018uwb}.

\subsection{Detectability of GWTC-3 events}\label{subsec:det}

 \begin{figure*}
\centering
 \includegraphics[width=\textwidth]{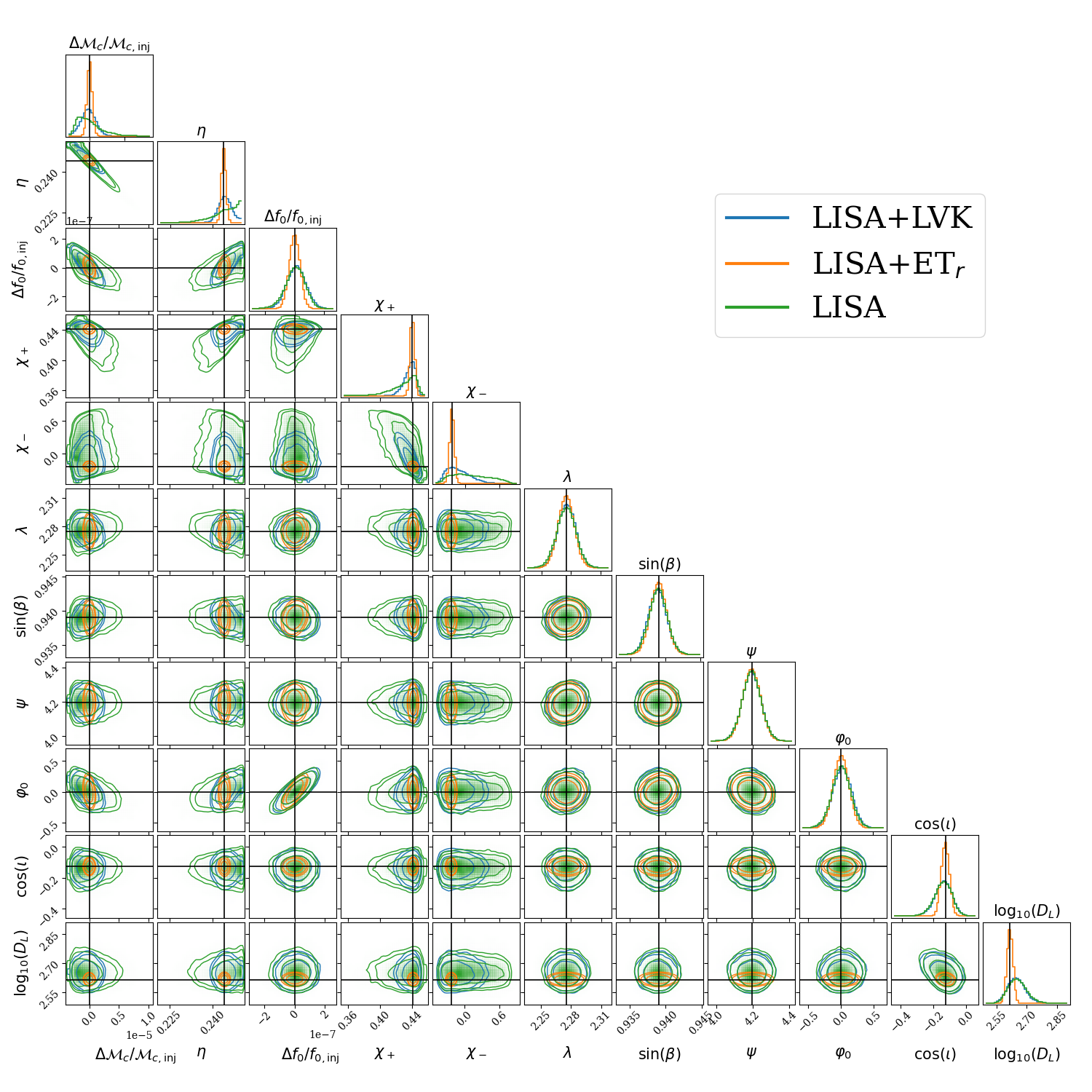}\\
 \centering
 \caption{Inferred posterior distribution for our GW191109$\_$010717-like system in the LISA only (green), LISA+LVK (blue) and LISA+ET$_r$ (orange) scenarios. The contours correspond to the 68, 90 and 95 $\%$ confidence regions. In the LISA+ET$_r$ case the covariance matrix of the Gaussian prior is computed from LVK samples and rescaled according to the improvement in SNR from LVK-like detectors to ET, as described in Sec.~\ref{sec:catalogue}. The sky location, polarization and initial phase distributions are the same in the three scenarios because we discard the information on those parameters coming the LVK data.}\label{fig:comp_configs_1}
\end{figure*}

As described in Sec.~\ref{sec:catalogue}, we use the masses, spins, distance and inclination posterior samples released in GWTC-3 and for each sample we compute the SNR in LISA randomizing over sky location, polarization and initial phase. Out of all the events in GWTC-3, only a few could be detected with LISA. We define the probability of a direct (archival) detection with LISA of a given event, $p_{\rm det}^8$ ($p_{\rm det}^5$), as the fraction of samples for this event that have SNR above 8 (5)\footnote{Our definition is similar to the standard one in literature, see e.g. Eq.~4 in \cite{Mandel:2018mve}, with the difference that we marginalize over the unknown event parameters using the LVK posterior. We then perform a Monte-Carlo integration and use the SNR as a detection statistic.}. Fig.~\ref{fig:snrs} displays the distribution of SNRs in LISA for the LVK events, assuming the CBE PSD. In each panel, we plot in color the events that have $p_{\rm det}^5>0.1$ when using the CBE PSD. As expected, more events pass this criterion as we observe the systems closer to merger and for a longer time (from left to right panel). In Table \ref{tab:pdets}, we give $p_{\rm det}^5$ and $p_{\rm det}^8$ for our three fiducial observational scenarios and for the events that have $p_{\rm det}^5>0.1$ in the most optimistic configuration ($t_c=10{\rm yr}$, $T_{\rm obs}=10{\rm yr}$, CBE PSD). These events are shown in color in the right panel of Fig.~\ref{fig:snrs}. The lower value is computed assuming the SciRDv1 PSD and the higher one assuming the CBE PSD, accounting for a $75\%$ duty cycle in both cases. We find that 13 events would have had a $10\%$ probability of being found in the LISA data stream through archival searches in that most optimistic case. If instead we use the SciRDv1 PSD this number drops to 6, and to 1 if in addition we observe the systems far from merger. In fact, only GW150914 would have been almost guaranteed to be observed with LISA, confirming once again how exceptionally loud this event was. Other than GW150914, only GW191109$\_$010717 would have had a higher than $10\%$ probability of being directly detected, and still, only in the most optimistic configuration.

\begin{table*}[hbtp!]
  \begin{center}
   \begin{tabular}{c *{4}{c|}}
   
   \cline{2-5}
   
   &  \multicolumn{3}{|c|}{GW191109$\_$010717} &  \multicolumn{1}{|c|}{GW200129$\_$065458} \\
   
    \cline{2-5} 
    
    &  \multicolumn{1}{|c|}{$t_c=10{\rm yr}$, $T_{\rm obs}=6{\rm yr}$} &
    
    \multicolumn{1}{|c|}{$t_c=6{\rm yr}$, $T_{\rm obs}=6{\rm yr}$} &
    
    \multicolumn{1}{|c|}{$t_c=10{\rm yr}$, $T_{\rm obs}=10{\rm yr}$} & 
    
    \multicolumn{1}{|c|}{$t_c=10{\rm yr}$, $T_{\rm obs}=10{\rm yr}$} \\
    
    \hline
    
    \multicolumn{1}{|c|}{$m_1 \ (M_{\odot})$} & \multicolumn{3}{|c|}{68} & \multicolumn{1}{|c|}{55} \\
    
    \hline
    
    \multicolumn{1}{|c|}{$m_2 \ (M_{\odot})$} & \multicolumn{3}{|c|}{63} & \multicolumn{1}{|c|}{27} \\
    
    \hline
    
    \multicolumn{1}{|c|}{$\chi_1$} & \multicolumn{3}{|c|}{-0.12} & \multicolumn{1}{|c|}{0.28} \\
    
    \hline

    \multicolumn{1}{|c|}{$\chi_2$} & \multicolumn{3}{|c|}{-0.41} & \multicolumn{1}{|c|}{-0.01} \\
    
    \hline
    
    \multicolumn{1}{|c|}{$\iota$} & \multicolumn{3}{|c|}{1.5} & \multicolumn{1}{|c|}{0.3} \\
    
    \hline
    
    \multicolumn{1}{|c|}{$D_L \ (\rm Mpc)$} & \multicolumn{3}{|c|}{327} & \multicolumn{1}{|c|}{1036} \\
    
    \hline
     
     \multicolumn{1}{|c|}{$f_0 \ (\rm mHz)$} & \multicolumn{1}{|c|}{7.8728} & \multicolumn{1}{|c|}{9.5336} & \multicolumn{1}{|c|}{7.8728} & \multicolumn{1}{|c|}{10.993} \\
    
    \hline    

    \multicolumn{1}{|c|}{$\lambda$} & \multicolumn{3}{|c|}{1.5} & \multicolumn{1}{|c|}{1.2} \\
    
    \hline
    
    \multicolumn{1}{|c|}{$\beta$} & \multicolumn{3}{|c|}{-0.3} & \multicolumn{1}{|c|}{-0.5} \\
    
    \hline 
    
    \multicolumn{1}{|c|}{$\psi$} & \multicolumn{3}{|c|}{1.0} & \multicolumn{1}{|c|}{1.9} \\
    
    \hline
    
     \multicolumn{1}{|c|}{$\phi$} & \multicolumn{3}{|c|}{0.0} & \multicolumn{1}{|c|}{0.0} \\
     
     \hline
     
     \multicolumn{1}{|c|}{SNR} & \multicolumn{1}{|c|}{7.7} & \multicolumn{1}{|c|}{10.2} & \multicolumn{1}{|c|}{11.9} & \multicolumn{1}{|c|}{5.1} \\
   
    \hline

   \end{tabular}
   \end{center}
    \caption{Parameters of the two events from GWTC-3 for which we perform parameter estimation. We consider the three time to coalescence and observation time combinations for GW191109$\_$010717. For GW191109$\_$010717 we consider only the most optimistic case, since it has not chances of being detected otherwise. The SNRs were computed using the SciRDv1 noise curve and not accounting for the duty cycle.}\label{tab:systems}
  \end{table*}

Assuming the events in GWTC-3 are independent, we can compute the probability that LISA would have detected at least one of them, directly and through archival searches\footnote{In practice, this is is given by $1-\prod_i(1-p_{\rm det,i})$, i.e. the complementary probability to detecting no event at all.}. We also compute the probability of detecting at least one event other than GW150914, which would have had high detection probability as we saw. Note that our computation assumes all events are observed with the same $t_c$ and $T_{\rm obs}$ combination, an unrealistic assumption, which leads to overly pessimistic or overly optimistic predictions. To bracket uncertainties, we can take the value obtained for $t_c=6{\rm yr}$, $T_{\rm obs}=10{\rm yr}$ (left column) as lower bound and the one for $t_c=10{\rm yr}$, $T_{\rm obs}=10{\rm yr}$ (right column) as higher bound. We find that with the CBE noise curve the probability of observing at least one event other than GW150914 is in the range $2.6 \%$-$29 \%$ for a direct detection, and $37\%$-$99 \%$ for an archival detection.

\subsection{Parameter estimation}\label{subsec:pe}

Due to its scientific importance, GW150914-like systems have often been used to assess the ability of LISA to observe SBHBs. Henceforth, we decide instead to focus on a GW191109$\_$010717-like system and perform parameter estimation for such system for the three $t_c$ and $T_{\rm obs}$ combinations. Moreover, we also perform parameter estimation for a GW200129$\_$065458-like system, which could be detected only through archival searches when $t_c=T_{\rm obs}=10{\rm yr}$. The events on which we perform parameter estimation lie rather in the tail of the high SNR distribution, which corresponds to the tail in the distance distribution. We further picked a sample that has representative masses (close to maximum {\it a posteriori}). For the parameter estimation (from now on) we use the SciRDv1 noise curve and do not account for the $75\%$ duty cycle. We give the parameters of the systems for which we perform parameter estimation together with their SNR in Table \ref{tab:systems}. 

\subsubsection{GW191109$\_$010717-like system}

 \begin{figure*}[hbtp!]
  \includegraphics[scale=0.37]{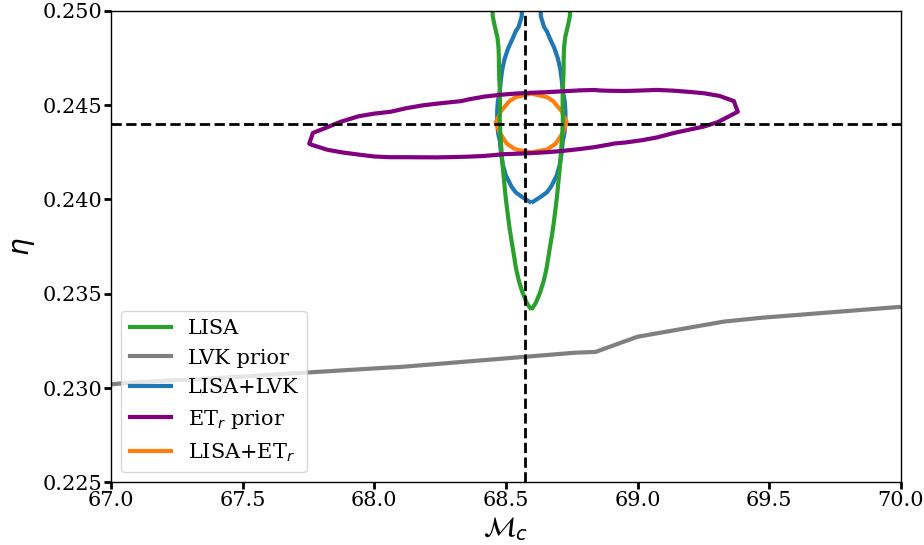}
    \includegraphics[scale=0.37]{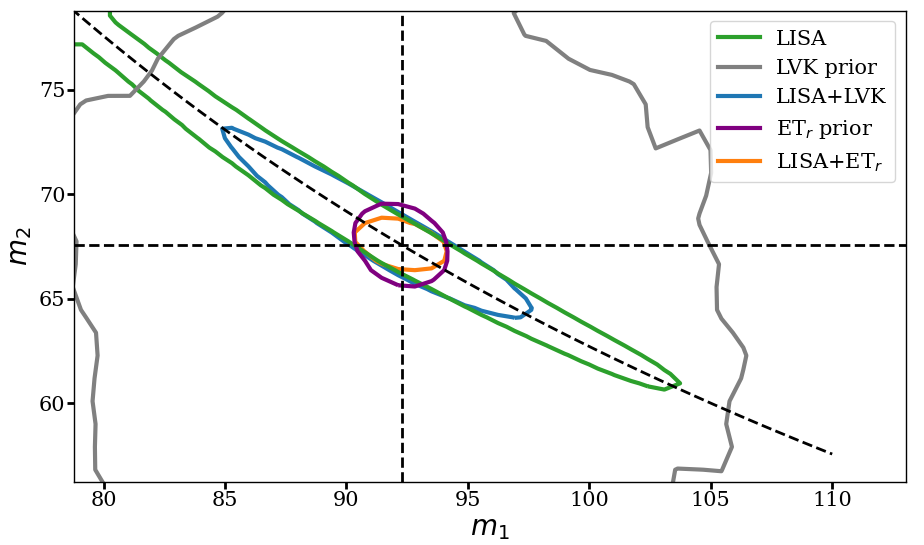}
   \caption{Left panel: $90\%$ contours for the chirp mass and the symmetric mass in different scenarios and for the different priors used. Right panel: same for the individual (redshifted) masses. Dashed lines indicate the injected values. On the right panel we also show the line in the ($m_1$,$m_2$) plane corresponding to the injected $\mathcal{M}_c$ value. As expected, LISA can measure the chirp mass very accurately, but not other mass combinations, leading to the \enquote{banana-like} correlation between $m_1$ and $m_2$. Thanks to this very accurate measurement, LISA can considerably (marginally) improve the measurement of $m_1$ and $m_2$ in the LISA+LVK (LISA+ET$_r$) scenario.}\label{fig:comp_masses_1_tobs10}  
 \end{figure*}

  \begin{figure*}[hbtp!]
  \includegraphics[scale=0.37]{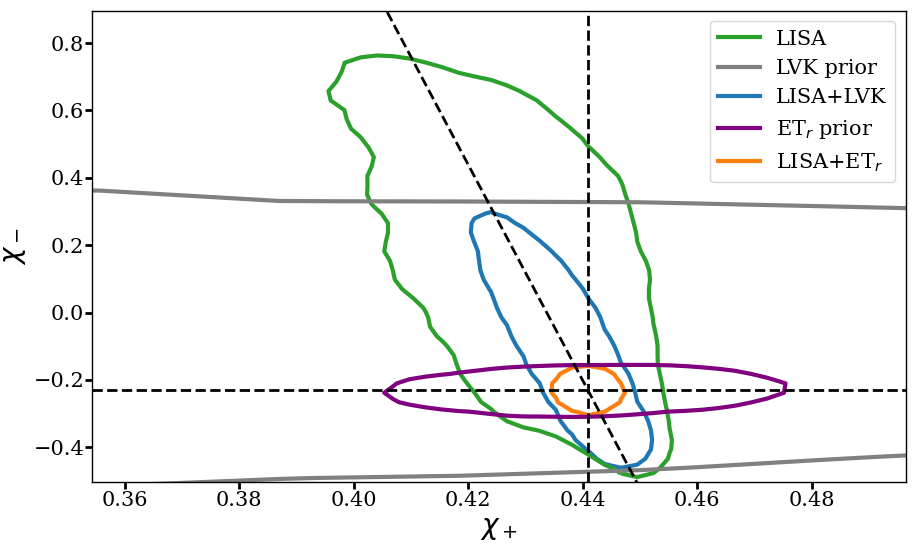}
    \includegraphics[scale=0.37]{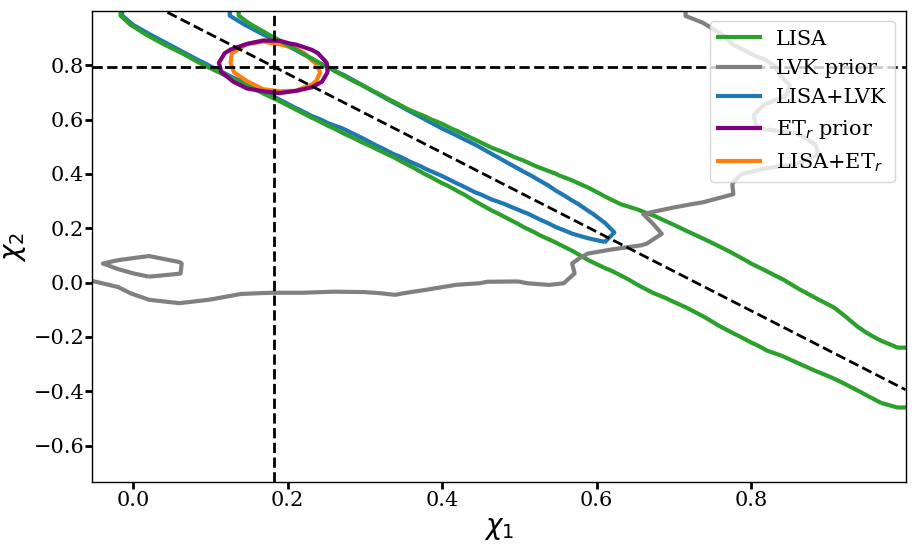}
   \caption{Left panel: $90\%$ contours for the symmetric and antisymmetric spin combinations in different scenarios and for the different priors used. Right panel: same for the individual spins. Dashed lines indicate the injected values. We also plot the lines in the ($\chi_+$,$\chi_-$) and ($\chi_1$,$\chi_2$) planes corresponding to the injected $\chi_{PN}$ value (see Eq.~\ref{eq:chi_pn}). LISA can measure $\chi_{PN}$ accurately, but not other spin combinations, leading to a strong correlation between $\chi_1$ and $\chi_2$. Thanks to this very accurate measurement, LISA can considerably (marginally) improve the measurement of $\chi_1$ and $\chi_2$ in the LISA+LVK (LISA+ET$_r$) scenario.}\label{fig:comp_spins_1_tobs10}  
 \end{figure*}

Figure \ref{fig:comp_configs_1} shows the posterior distribution for GW191109$\_$010717-like system in the $t_c=T_{\rm obs}=10{\rm yr}$ case, in the form of a {\it corner plot}\cite{corner}. There, we compare the distribution obtained with LISA alone to the one obtained in the LISA+LVK and LISA+ET$_r$ scenarios. For sake of clarity we show the distribution of the relative error in chirp mass and initial frequency, where $\mathcal{M}_{c, {\rm inj}}$ and $f_{0, {\rm inj}}$ are the values of the injected system. We refer to \cite{Toubiana:2020cqv} for a discussion on the measurement accuracy and correlation between parameters in the LISA posterior, and emphasize instead the difference between the three scenarios. 
Because we retain no information for the LVK data on sky position, polarization and phase, the distribution of these parameters is the same in the three scenarios. As discussed in \cite{Toubiana:2020cqv}, we find that LISA provides a very precise sky localization, typically within $1 {\rm deg}^2$. 

We zoom on the mass parameters in Fig.~\ref{fig:comp_masses_1_tobs10}. In the left panel we show the $90\%$ contours for $\mathcal{M}_c$ and $\eta$ in the three scenarios, as well as for the LVK and {ET}$_r$ priors. The black dashed lines indicate the injected values. It is clear that LISA provides the best chirp mass measurement, even better than ET$_r$, despite a modest SNR of 11.9. This is a result of LISA observing many GW cycles during the early inspiral. In that regime, the phase evolution is dictated by the leading order post-Newtonian term, which depends only on $\mathcal{M}_c$ \cite{Blanchet:2013haa}. We also find that in this favorable case, where we observe the system chirping and for a long time, LISA can actually provide a better estimate of $\eta$ than current ground-based detectors. As a consequence, the determination of the individual masses is much tighter when combining LISA with LVK-like observations, as can be seen on the right panel. However, ET$_r$ can constrain the symmetric mass ratio much better than LISA, and, as a consequence, the measurement of the individual masses improves only marginally when including LISA information. 

The determination of spins follows a similar pattern, as can be seen in Fig.~\ref{fig:comp_spins_1_tobs10}. The symmetric spin combination $\chi_+$ is better measured with LISA than with ground-based detectors, but ET$_r$ provides a much better constraint on the antisymmetric spin combination. Therefore, the individual spins are much better constrained in the LISA+LVK scenario relative to the LVK-based prior, but LISA improves only marginally the determination of individual spins relative to ET$_r$. Note that in fact the spin combination that LISA can the most precisely measure is \cite{Toubiana:2020cqv}
\begin{equation}\label{eq:chi_pn}
  \chi_{PN}=\frac{\eta}{113} \left( (113 q+75) \chi_1+ (113 q^{-1}+75) \chi_2 \right). 
\end{equation}
We plot in dashed lines the injected value of $\chi_{PN}$ in both panels. In the nearly equal mass case, $\chi_{\rm PN} \simeq \frac{94}{113}\chi_+$, explaining why $\chi_+$ is measured that well with LISA.    

Figure \ref{fig:comp_prior_1_tobs10} in Appendix \ref{app:plots} shows the full comparison between the posteriors obtained with our MCMC code and the priors used in the LISA+ET$_r$ scenario. Moreover, Fig.~\ref{fig:comp_prior_1_tobs6} shows the same comparison in the LISA+ET$_r$ scenario for the $t_c=10{\rm yr}$, $T_{\rm obs}=6{\rm yr}$ case. So far from merger, LISA can no longer constrain $\eta$ and $\chi_+$, since it does not observe the system chirping enough. However, it can still measure $\mathcal{M}_c$ better than ET$_r$, and as a consequence this improves the measurement of parameters that correlate with the chirp mass, such as the symmetric mass ratio and the effective spin. From these figures, we can also conclude that the determination of the distance and the inclination in the LISA+ET$_r$ scenario seen in Fig.~\ref{fig:comp_configs_1} comes from the ET$_r$ measurement alone.

  \begin{figure*}
\centering
 \includegraphics[width=\textwidth]{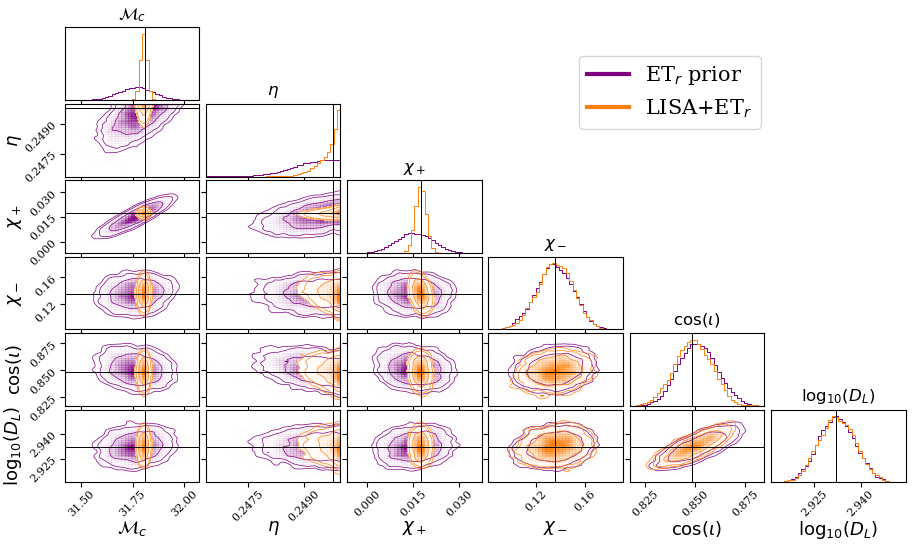}\\
 \centering
 \caption{Comparison between the prior used in the LISA+ET$_r$ scenario and the parameter estimation results for GW200129$\_$065458. Even in this low SNR case, LISA can provide more information about the chirp mass than ET$_r$, thanks to the large number of cycles in band. On the other hand, all the information on inclination and distance comes from the ET$_r$ prior.}\label{fig:comp_prior_2_et}
\end{figure*}

 \subsubsection{GW200129$\_$065458-like system}
 
\begin{figure*}
\centering
 \includegraphics[width=\textwidth]{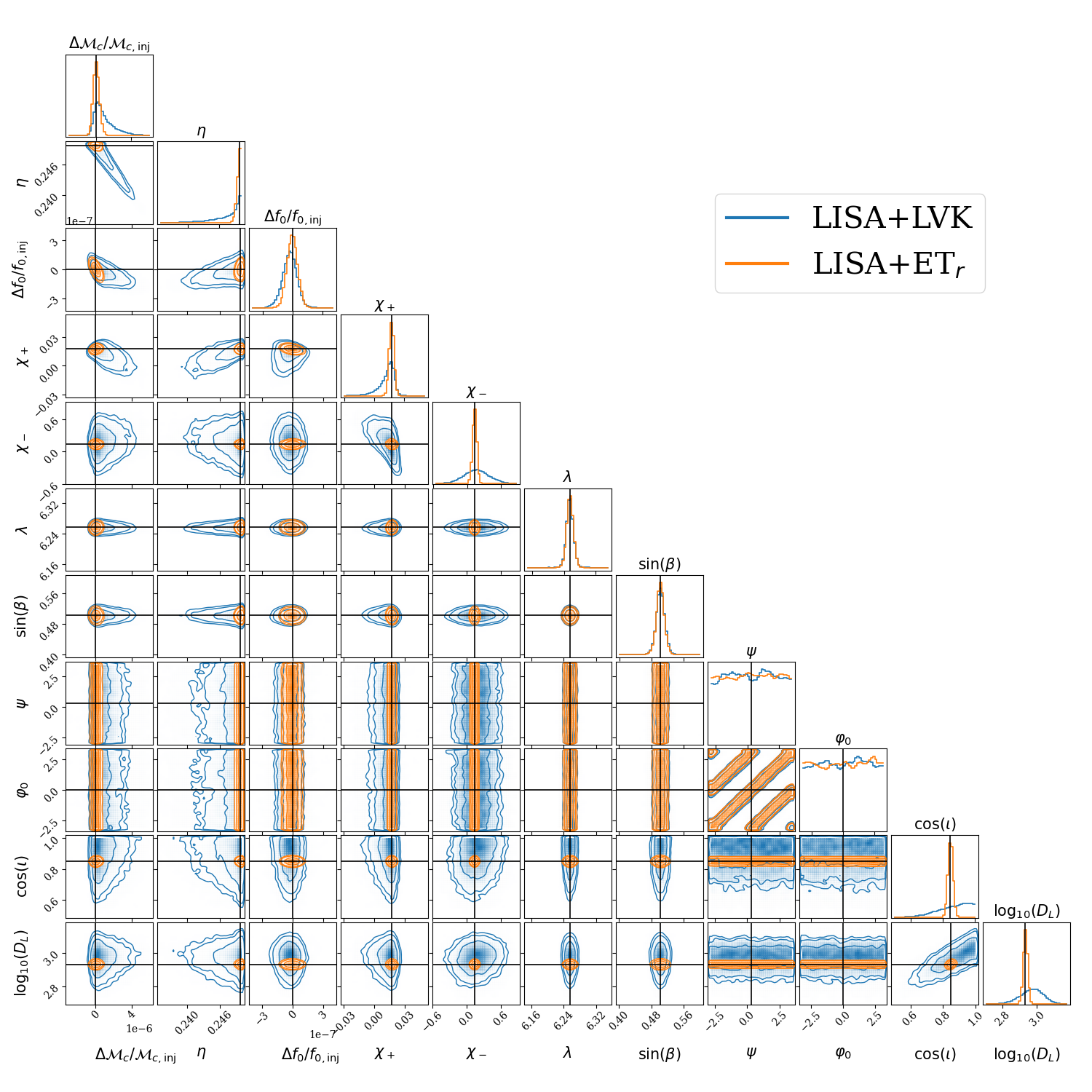}\\
 \centering
 \caption{Inferred posterior distribution for GW200129$\_$065458 in the LISA+LVK (blue) and LISA+ET$_r$ (orange) scenarios. }\label{fig:comp_configs_2}
\end{figure*}

 The posterior distribution for GW200129$\_$065458-like system in both the LISA+LVK and LISA+ET$_r$ scenarios are shown in Fig.~\ref{fig:comp_configs_2}. The sky localization is a bit worse for this system, around $2{\rm deg}^2$, due to its very low SNR. The benefit of using LISA can be better seen in Fig.~\ref{fig:comp_prior_2_et}, where we compare the posterior distribution obtained in the LISA+ET$_r$ scenario to the prior. As in the case of GW191109$\_$010717-like system, LISA allows a much more precise determination of the chirp mass. This in turn provides a better measurement of parameters that correlate with $\mathcal{M}_c$, such as the symmetric mass ratio and the effective spin. When transforming to individual parameters we find that the determination of the masses improves a bit when including LISA, while the determination of spins improves only marginally, as can be seen in Figs.~\ref{fig:comp_masses_2_tobs10} and \ref{fig:comp_spins_2_tobs10}.

 \section{Conclusion}\label{sec:ccl}
 
 Our goal in this paper was to realistically assess the ability of LISA to detect SBHBs. To do so, we have estimated how many events from GWTC-3 could be detected by LISA using the publicly released data. We considered the possibility of detecting these systems with LISA before ground-based detectors as well as through archival searches, using posteriors for the sources detected by the LVK as priors in the LISA search. We found that even in the most optimistic configuration, only 13 events would have had a higher than $10 \%$ probability of being detected, and most of them only through archival searches. Only the golden event GW150914 would have been almost certainly detected. The probability of detecting any other event from GWTC-3 is in the range $2.6$ \%-$29 \%$ for a direct detection, and $37 \%$-$99 \%$ for an archival detection for a realistic LISA noise curve.
 
 We have justified our choice of SNR=8 as threshold for direct detection and SNR=5 for archival detection by examining the behavior of MCMC chains for near threshold systems. Below these values the chain can \enquote{lose} the signal, implying that the detection is not statistically sound. Note that our threshold for the archival detection is slightly higher than what was previously suggested (4) \cite{Wong:2018uwb}.  
 
 Next, we performed parameter estimation for a GW191109$\_$010717-like system that passes the threshold for direct detection with LISA and a GW200129$\_$065458-like system, which could be detected through the archival search. We created mock multiband observations together with LVK-like detectors by imposing a Gaussian prior on the source parameters with covariance matrix computed from the samples released in GWTC-3. We also mimic a LISA+ET$_r$ case by rescaling the covariance matrix according to the improvement in SNR from LVK to ET. We found that even for very low SNR systems, LISA can better measure the mass than ET$_r$ thanks to observing the binary for many orbital cycles. This leads to a better measurement also of parameters that correlate with the chirp mass, such as the mass ratio and the effective spin, when adding LISA information relative to ground-based detectors alone. However, when translating into individual masses and spins, the improvement is often marginal. Thus, the information provided by LISA could be helpful in distinguishing between astrophysical models that predict distinctive features in the chirp mass and effective spin distributions. Moreover, for systems loud enough to be directly detected, LISA would localize the sources typically within $1{\rm deg}^2$, which would facilitate the search for a possible electromagnetic counterpart, in particular merger counterparts. Although we did not explore this possibility here, we expect that such multiband observations should improve constraints on low-frequency modifications to the GW signal, which could arise due to astrophysical or modified gravity effects. 
 
Our study provides a realistic estimate of the ability of LISA to detect SBHBs and constrain their parameters. Note that as the total mass of the system increases, the LISA horizon expands while the LVK one shrinks, until they cross at a few $100 M_{\odot}$. Therefore, intermediate mass black hole binaries, an interesting class of sources for LISA \cite{Miller:2008fi,Jani:2019ffg}, might be underrepresented in our study. Finally, here we assumed all binaries to be quasicircular and nonprecessing. While the second assumption has little impact for LISA (though it does play a role for multiband observations), it will be necessary to include eccentricity in future studies. Important steps have been done recently in this direction in \cite{Buscicchio:2021dph,Klein:2022rbf}.

\begin{acknowledgments}
S.B. and S.M. acknowledge support from the French space agency CNES in the framework of LISA. A.T. is thankful to Jonathan Gair for valuable input in the preparation of this work. 
 \end{acknowledgments}

    \begin{figure*}[htbp!]
        \includegraphics[width=\textwidth]{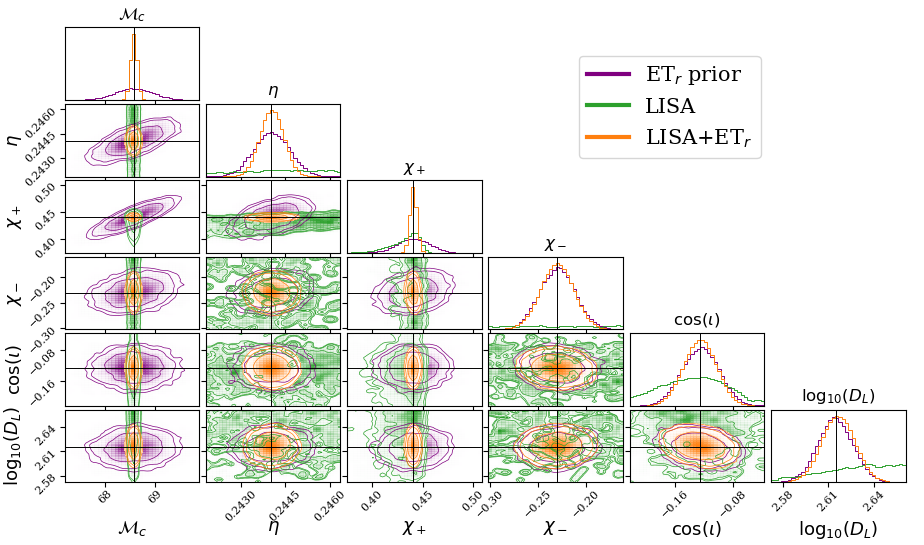}
   \caption{Comparison between the prior used in the LISA+LVK (upper) and LISA+ET$_r$ (lower) scenarios and the parameter estimation results, with and without the ground-based informed prior, for GW191109$\_$010717 in the $t_c=T_{\rm obs}=10{\rm yr}$.}\label{fig:comp_prior_1_tobs10}  
 \end{figure*}
 
 \appendix
 
 \section{Complementary plots}\label{app:plots}

We provide here the remaining plots that are commented in the main text.

   \begin{figure*}[hbtp!]
 \includegraphics[width=\textwidth]{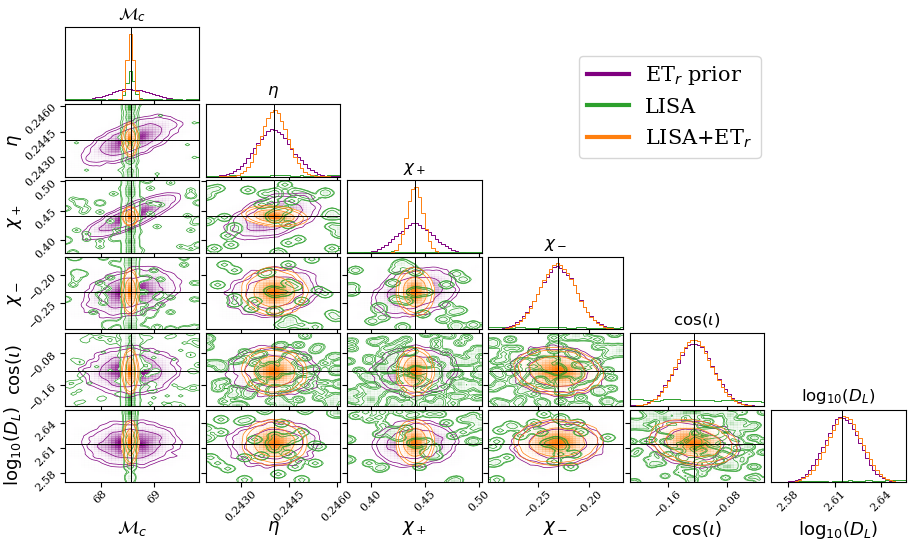}\\
 \centering
 \caption{Same as Fig.~\ref{fig:comp_prior_1_tobs10} for $T_{\rm obs}=6{\rm yr}$.}\label{fig:comp_prior_1_tobs6}
\end{figure*}

 \begin{figure*}[hbtp!]
  \includegraphics[scale=0.37]{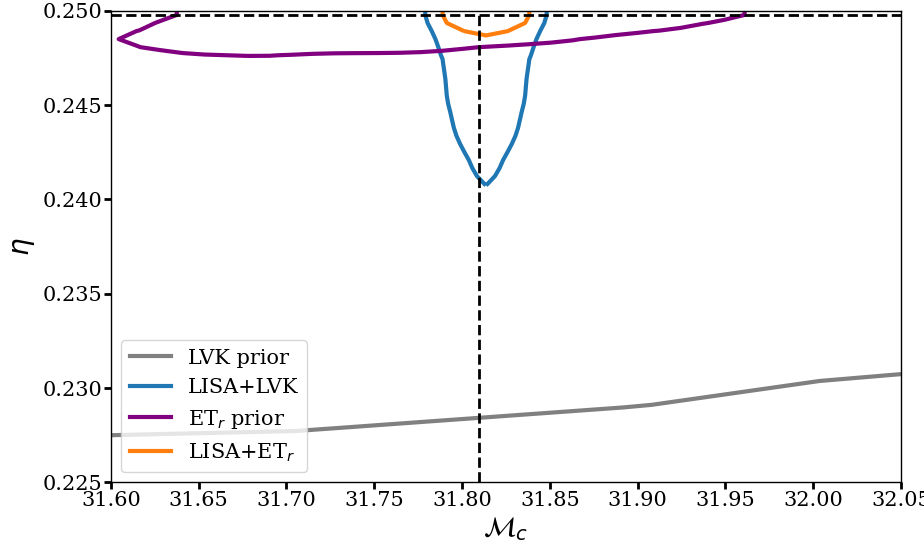}
    \includegraphics[scale=0.37]{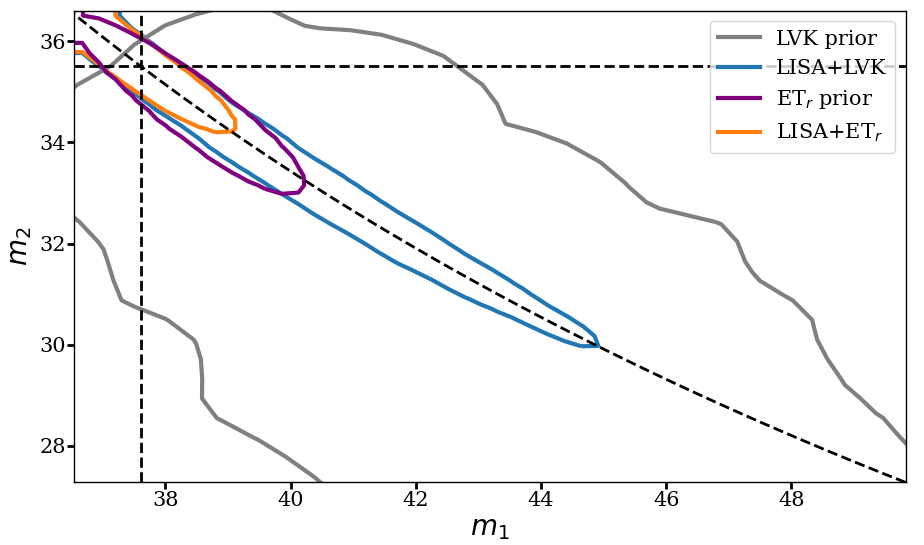}
   \caption{Sames as Fig.~\ref{fig:comp_masses_1_tobs10} for our GW200129$\_$065458-like system. Note the posterior is truncated by the prior boundary at $\eta=\frac{1}{4}$.} Including LISA has a greater impact on the measurement of individual masses than for our GW191109$\_$010717-like system.\label{fig:comp_masses_2_tobs10}  
 \end{figure*}
 \vspace{0.00mm}
 \begin{figure*}[hbtp!]
  \includegraphics[scale=0.37]{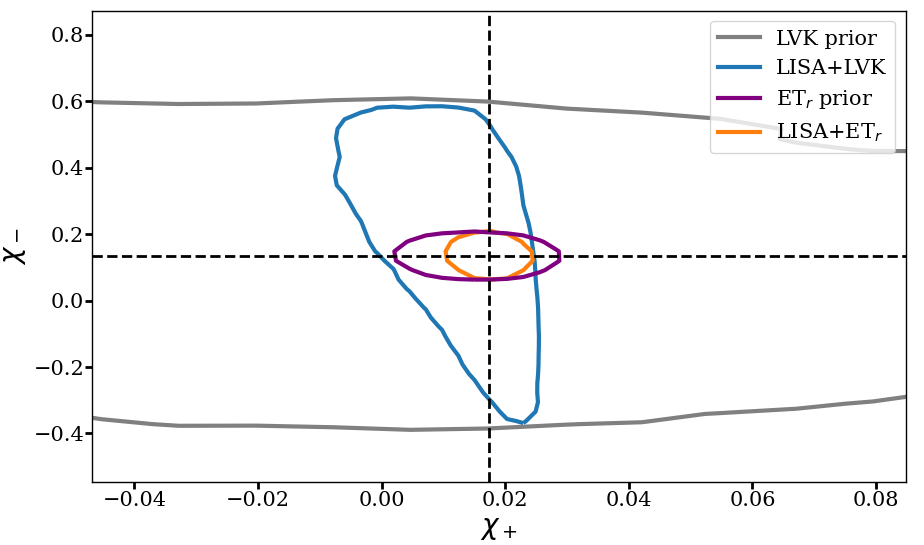}
    \includegraphics[scale=0.37]{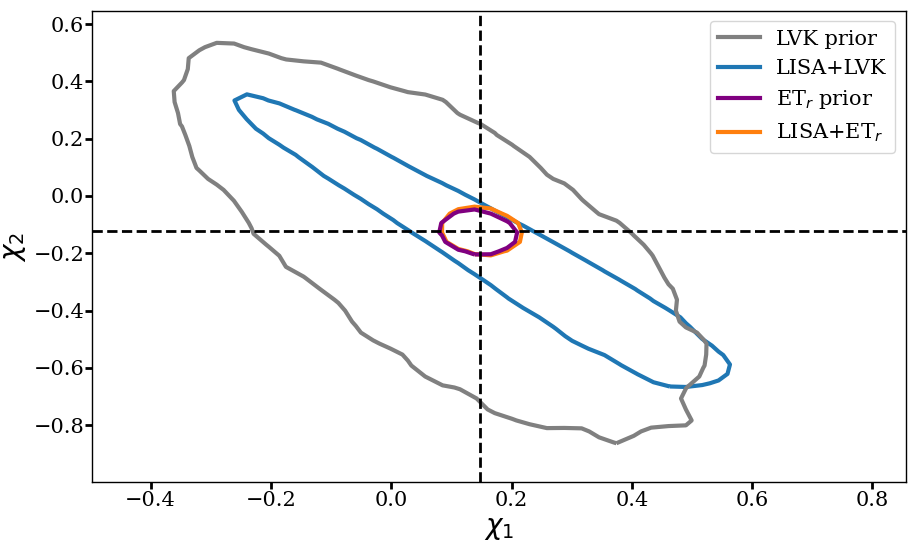}
   \caption{Sames as Fig.~\ref{fig:comp_spins_1_tobs10} for our GW200129$\_$065458-like system. As for our GW191109$\_$010717-like system, although LISA can help constraining the effective spin, the impact on the measurement of individual spins in marginal.  }\label{fig:comp_spins_2_tobs10}  
 \end{figure*}

  \FloatBarrier
\bibliography{Ref}
\end{document}